\documentclass[%
 reprint,
 amsmath,amssymb,
 aps,
]{revtex4-2}

\usepackage{comment}
\usepackage{graphicx}
\usepackage{dcolumn}
\usepackage{bm}

\begin{document}

\preprint{APS/123-QED}

\title{Multiple memory formation in glassy landscapes}

\author{Chloe W. Lindeman}
 \email{cwlindeman@uchicago.edu}
\author{Sidney R. Nagel}%
\affiliation{%
 Department of Physics and The James Franck and Enrico Fermi Institutes \\ 
 University  of  Chicago, \\ 
 Chicago,  IL  60637,  USA.
}%

\date{\today}

\begin{abstract}
Cyclically sheared jammed packings form memories of the shear amplitude at which they were trained by falling into periodic orbits where each particle returns to the identical position in subsequent cycles. While simple models that treat clusters of rearranging particles as isolated two-state systems offer insight into this memory formation, they fail to account for the long training times and multi-period orbits observed in simulated sheared packings. We show that adding interactions between rearranging clusters overcomes these deficiencies.  
In addition, interactions allow simultaneous encoding of multiple memories which would not have been possible otherwise. These memories are different in an essential way from those found in other systems, such as multiple transient memories observed in sheared suspensions, and contain information about the strength of the interactions.
\end{abstract}

\maketitle


\section{\label{sec:level1}Introduction}

Memories in matter can be created in a multitude of ways~\cite{KeimPaulsen2019}. Of interest here is a particular form of memory in which a jammed packing of particles subjected to training by a cyclic quasistatic shear can form a memory of the amplitude at which the shear was applied~\cite{fiocco2014encoding,Keim14,Royer15,lavrentovich2017period, Adhikari18,Sood19}. The memory is encoded because the system falls into a periodic orbit in which the packing revisits precisely the same states during each applied shear cycle. The periodic orbit is disturbed if the shear amplitude is altered so that a memory of the training amplitude can be ``read out’’ by tracking the particle displacements after cycles of increasing strain. 

This memory formation is reminiscent of the ones found in suspensions of non-Brownian suspensions~\cite{Pine05,Corte08,Keim11,Keim13b,Paulsen14}. However, for such suspensions each particle only interacts if it collides with a neighbor; in jammed particle packings, the particles are in enduring contact with their neighbors throughout each cycle. In this case it is less clear how the memory is formed. The energy landscapes are different: the jammed systems have an exponential number of well-defined local energy minima, while the suspensions have very large flat-bottomed ground states. In the case of suspensions, the orbit is reversible along each cycle and there are no energy barriers that need to be overcome.  In contrast, jammed systems deform via rearrangements between clusters of particles that occur in one direction of the shear and that undo themselves at a different amplitude as the system is sheared in the reverse direction~\cite{fiocco2014encoding}. The motion is thus not reversible within a cycle but is still periodic.  Because jammed packings exist in a very rugged, high-dimensional and complex energy landscape~\cite{Liu10}, it is astonishing that these systems can find a periodic orbit at all and moreover that the periodic orbit can be discovered relatively rapidly.  

Aspects of the periodic memories encoded in jammed packings 
have been modeled in a variety of ways~\cite{Ortin91,fiocco2014encoding,Fiocco15,Paulsen19,Mungan19}.  
We consider here a model that was motivated by the existence of localized regions in a disordered material which are particularly prone to rearrangements due to applied external forcing~\cite{Ortin91}. 
This is based on the Preisach model~\cite{Preisach35} originally proposed for magnetic systems. This model considers independent, non-interacting defects, each of which can exist in either of two states with an energy barrier between them. Due to an applied external strain, a defect can flip between the two states; however, the strain to flip in the ``forward'' direction is not necessarily the same as for it to flip in the opposite ``backwards'' direction. Each defect, known as a $\textit{hysteron}$, is thus an elementary unit of hysteresis in the system.  While Preisach models have been successful at describing many aspects of the memories, there are certain phenomena that they do not capture at all such as long training times and sub-harmonic response.

In this paper, we generalize this type of model by including interactions between hysterons so that the applied strains for one hysteron to flip between its two states depends on the state of the others.  As we will show, this generalization not only exhibits some of the phenomena not possible without interactions but also leads to a new type of memory that, to our knowledge, has not yet been identified in other systems.

\section*{Model}

\subsection*{Non-interacting model --- successes and failures}
\begin{figure*}
\centering
\includegraphics[width=17.8cm]{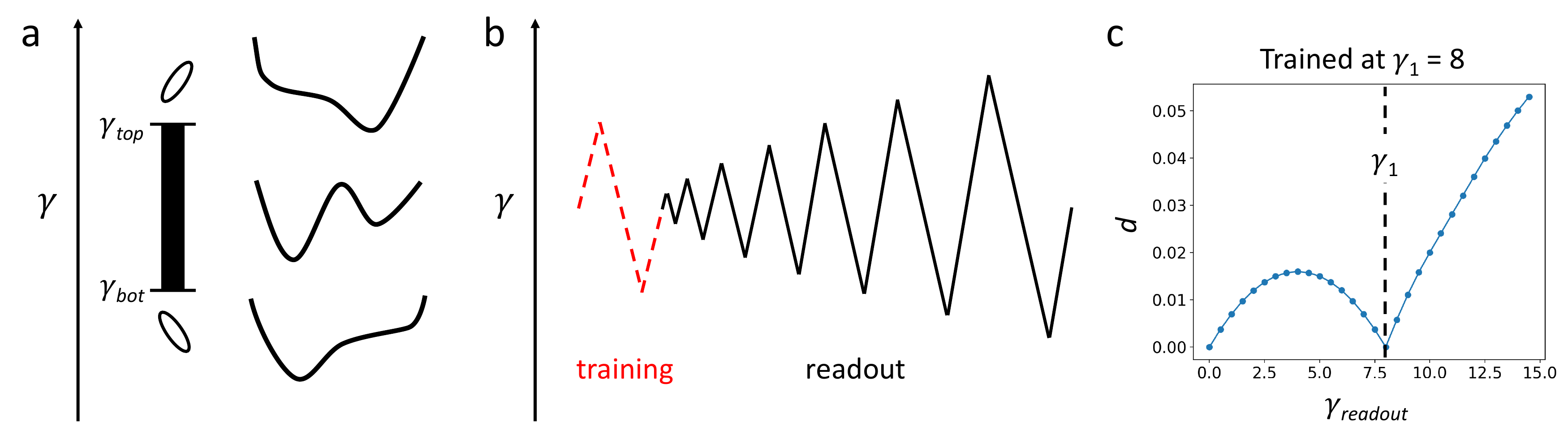}
\caption{(a) A single hysteron with flipping strains $\gamma_{top}$ and $\gamma_{bot}$.  The configuration can be modeled as a double-well potential as shown.  For $\gamma > \gamma_{top}$ the hysteron is forced into the / state (\textit{i.e.} the right-most well), and for $\gamma < \gamma_{bot}$ the hysteron is forced into the \textbackslash \ state (\textit{i.e.} the left-most well).  For intermediate values of the strain, $\gamma_{bot} < \gamma < \gamma_{top}$, the 
hysteron can reside in either state as prescribed by its preparation history.  (b) Example of training (dashed red) and readout (solid black) cycles. After training, the state of each hysteron is recorded; this state
is compared with the state after each readout cycle and the fractional difference is recorded
as $d$.  (c) An example of $d$ versus the readout 
strain, $\gamma_{readout}$, for a system of independent hysterons trained at $\gamma_1 = 8$.
}
\label{hysteron}
\end{figure*}

Consider an ensemble of hysterons 
in the presence of an externally applied shear $\gamma$. 
We denote the two possible states of a hysteron by \textbackslash \  and /.  Its current state is determined by its previous state (that is, its history) and the current 
value of the shear. In terms of the energy landscape, this is equivalent to having a double-well
potential such as the one shown in Fig.~\ref{hysteron}a. Any given hysteron is fully described
by its flipping strains $\gamma_{top}$ and $\gamma_{bot}$. For a system with a broad distribution of hysteron parameters, one can train the system as one would a jammed packing, with shear cycles of type 
$(0 \rightarrow \gamma \rightarrow 0 \rightarrow -\gamma \rightarrow 0)$ as in 
\cite{fiocco2014encoding, lavrentovich2017period,mungan2019networks} and as shown by the 
red (dashed) saw-tooth curve in Fig.~\ref{hysteron}b. The readout protocol is illustrated by the black (solid) saw-tooth curve which, starting at small amplitude, has an increasing amplitude for each subsequent cycle. The amplitude of the largest previously applied 
strain can be read out by measuring $d$, the fraction of hysterons that have changed 
their state at the end of each readout cycle. 
There is a sharp cusp in $d$ when the readout amplitude equals the training amplitude, as shown in Fig.~\ref{hysteron}c.  

The Preisach model gives rise to a special type of memory called $\textit{return-point memory}$,
where the system remembers extremal values of the applied shear~\cite{sethna1993hysteresis}. Return-point memory has the property that memories can be stored only in a particular order; an applied strain will be erased as soon as another larger strain is applied whereas if a smaller strain is applied the previously encoded memories of larger strains remain. 
The Preisach model and return-point memory in general have been 
applied to a variety of jammed systems, from simulations of binary mixtures of 
interacting particles~\cite{fiocco2014encoding} to experiments of sheared 2D amorphous solids~\cite{keim2020global}. This simple
model captures many aspects of jammed systems; in particular, it 
produces cusp-like memories at the amplitude where the training was applied. However, it fails to describe the complexity both in the training and in the periodic orbits themselves. 

The number of driving cycles $\tau$ that a physical system
takes to reach a periodic orbit is an important parameter. For small shear 
amplitudes, $\tau$ can be just a few cycles. As the shear amplitude grows, however, $\tau$ becomes large and appears to diverge at some critical amplitude~\cite{lavrentovich2017period,regev2013onset}. In the Preisach model, however, $\tau \le 1$; this behavior can be understood by noting that each individual hysteron 
requires at most one cycle before it reaches
a periodic orbit.  This represents
a significant discrepancy with simulations of jammed packings. 

A second important aspect of the memory formation is the period $T$ of the orbit relative to the driving cycle. Simulations of frictional grains have shown that it is possible to fall into orbits that take many driving cycles for the particles to return to their original positions~\cite{Royer15}.  For frictionless jammed packings, the period $T$ grows for systems near the jamming transition~\cite{lavrentovich2017period}. In the Preisach model, $T = 1$ which can again be understood by considering a single hysteron under cyclic shear.

\subsection*{Including interactions between hysterons}
Work on hysteron-based models 
has primarily focused on how cyclically sheared jammed systems can result in return-point memory behavior.
Here, we present a modification to the Preisach model that includes interactions between hysterons. We will show how these interactions can 
explain the observed existence of $\tau > 1$ and $T > 1$ --- behavior not possible in the non-interacting model. 
In addition, we show that interactions produce a novel form of memory that allows the recall of training amplitudes that are smaller than those subsequently applied.  This new memory is not stored locally in the output (that is, not in a cusp appearing at the training strain as seen for example in Fig.~\ref{hysteron}c) but in the overall amplitude of the response. 

We introduce interactions between hysterons by making the flipping strains of hysteron $i$ depend on the orientations of its $n$ interacting neighbors. Since each neighbor could be in the \textbackslash \  or the / orientation, there are $2^n$ possible microstates, each of which induces a different value for the top (and likewise the bottom) flipping strain of hysteron $i$. How each microstate determines the flipping strains of hysteron $i$ can be chosen in a variety of ways. For simplicity we report here the choice that each microstate produces an  uncorrelated random additive shift to the non-interacting value of the flipping strain:
$\gamma_{i,top} = \gamma^0_{i,top}+\Delta_{i,top}^{nn}$, 
where $\gamma^0_{i,top}$ is the value of the top flipping strain without interactions and $\Delta_{i,top}^{nn}$ is the shift to that value due to the the particular microstate ${nn}$ of its neighbors. Similar rules were chosen for determining $\gamma_{i,bot}$.
We have also studied other possible rules, such as having each pairwise interaction be independent of the configuration of the other hysterons in the microstate.  The results that we report here are qualitatively insensitive to the choice.
 
In order to implement this model, we must choose the total number $N$ of hysterons in the system and the distributions of [$\gamma^0_{i,top}$, $\gamma^0_{i,bot}$, $\Delta^{nn}_{i,top}$, $\Delta^{nn}_{i,bot}$].  We also need to specify the rules that indicate which  hysterons interact with one another.
We have investigated two possible sets of interactions: (i) a \textit{mean-field model}, in which all $N$ hysterons interact;
(ii) a \textit{1D model}, in which we order the $N$ hysterons on a line (with periodic boundary conditions)
and add interactions between each hysteron and its $L$ neighbors on either side. We measured memory effects in the 1D model for large systems $N>>L$ 
and found them to be indistinguishable in both shape and magnitude from our results for 
the mean-field simulations as long as each hysteron interacts with the same number of neighbors (\textit{i.e.}, the size of the mean-field system, $N_{MF}$, is chosen to be the same as $2L+1$ in the 1D model). 

To determine the initial state of the system, a random sequence of $N$ \textbackslash \  and
/ states is chosen and the system is ``relaxed'' so that any unstable hysterons are stabilized.
During a shear cycle, the hysterons are flipped one at a time and the system is relaxed after each step. For each configuration, $\tau$ and $T$ are recorded.

We follow standard training and readout protocols~\cite{fiocco2014encoding, lavrentovich2017period}. We apply one or more training cycles and
record the state of each hysteron; this is our reference state. To read out, we apply cycles of increasing amplitude and record the fractional difference $d$ between the current state and the reference state after
each readout cycle. A typical plot of $d$ versus $\gamma_{readout}$ for a 
non-interacting system of hysterons is shown in
Fig.~\ref{hysteron}c.

To explore the memory capacity of this model, we train systems with a protocol involving two amplitudes: first, we fully train the system at a shear $\gamma_1$ (that is, apply as many training cycles as needed so that the system is in a periodic state at that amplitude of shear), then we apply a \textit{single} cycle at a second shear amplitude $\gamma_2$.  In a non-interacting model, if $\gamma_2 > \gamma_1$, the second amplitude would immediately (\textit{i.e.}, within a single cycle) erase any memory of of the first input; in that case, two copies of the same system trained at different values of $\gamma_1$ but with the same value of the larger amplitude $\gamma_2$ would look identical. When interactions are added, a second larger-amplitude shear cycle does \textit{not} necessarily immediately erase the memory of the smaller, previously stored, input amplitude.  Thus it is possible to observe more complex behavior in the storage of multiple memories in this model. 

\section*{Results from simulations}

In order to determine the role of interactions, we simulate the model for different values of $N$ and with distributions of [$\gamma^0_{i,top}$, $\gamma^0_{i,bot}$, $\Delta^{nn}_{i,top}$, $\Delta^{nn}_{i,bot}$] as described in the Methods section below.  The qualitative behavior of the model does not depend critically on the choices mode for these distributions as long as the distributions are sufficiently broad.  

\begin{figure}
\centering
\centerline{\includegraphics[width=8.7cm]{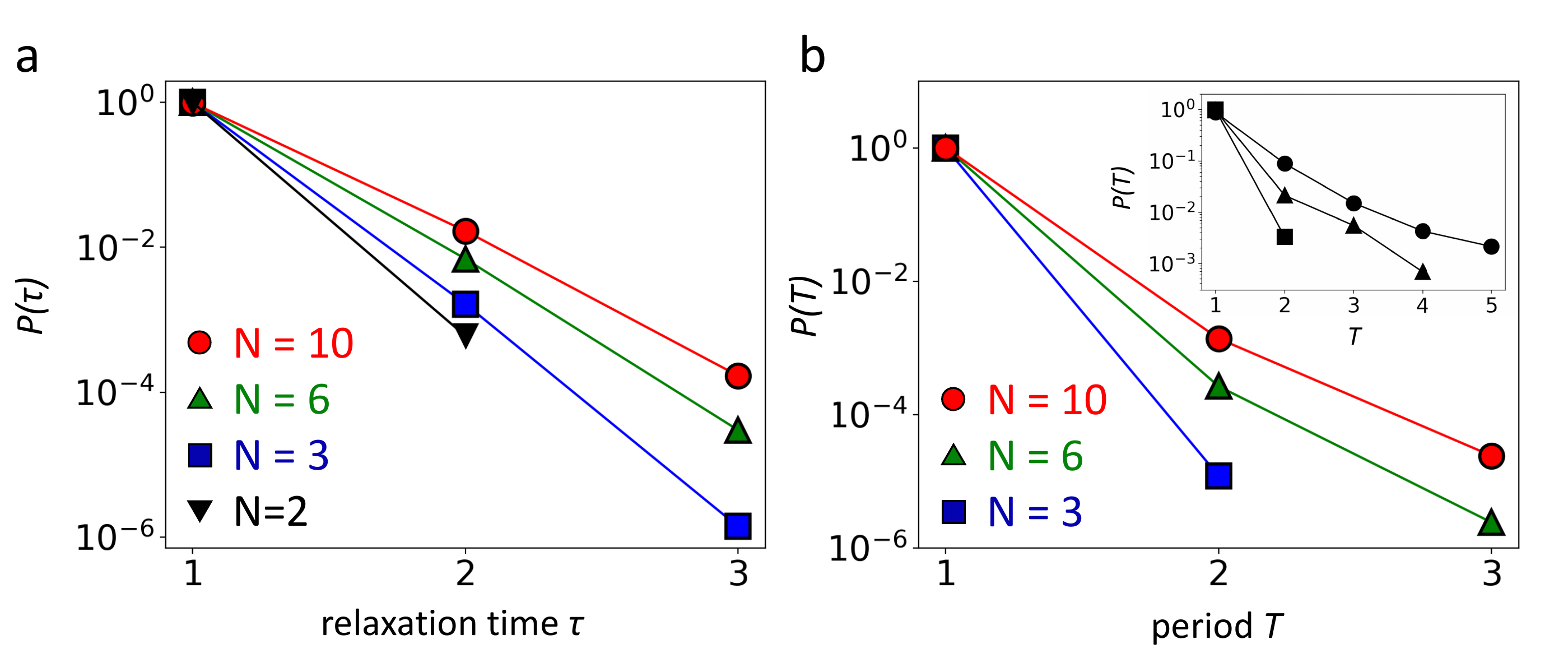}}
\caption{Probability distributions of (a) $\tau$ and (b) $T$ in a mean-field model of interacting hysterons where the parameter $A$ in the distribution of interaction strengths defined in the Methods section was chosen to be $A = 0.5$. Inset in (b) shows $P(T)$ versus $T$ from simulations of cyclically 
sheared jammed packings, adapted from~\cite{lavrentovich2017period}. In the inset, higher probability
curves correspond to lower pressure where the system is closer to the jamming threshold.}
\label{T-and-tau}
\end{figure}

We find that interactions allow for both $\tau$ and $T$ values greater than $1$. Figures~\ref{T-and-tau}a and b show the probabilities $P(\tau)$ and $P(T)$ for finding a configuration with a training time $\tau$ or period $T$  
for different system sizes in the mean-field model. 
In both cases, the probability decays approximately exponentially with $\tau$ or $T$ and increases as the number of interacting hysterons is increased. 
It is possible to get $\tau = 2$ for systems as small as $N=2$ interacting hysterons and to get $T = 2$ for systems as small as $N=3$. 

The inset of Fig.~\ref{T-and-tau}b
shows $P(T)$ versus $T$ for simulations of jammed systems cyclically sheared 
in the quasi-static limit~\cite{lavrentovich2017period}.
In those simulations, $P(T)$ also decayed approximately exponentially with $T$ and increased as the system was brought closer to the jamming transition where the range of elastic interactions increases.  This is consistent with our mean-field model where $P(T)$ increases with the number of interacting neighbors $N$.

\begin{figure*}
\includegraphics[width=17.8cm]{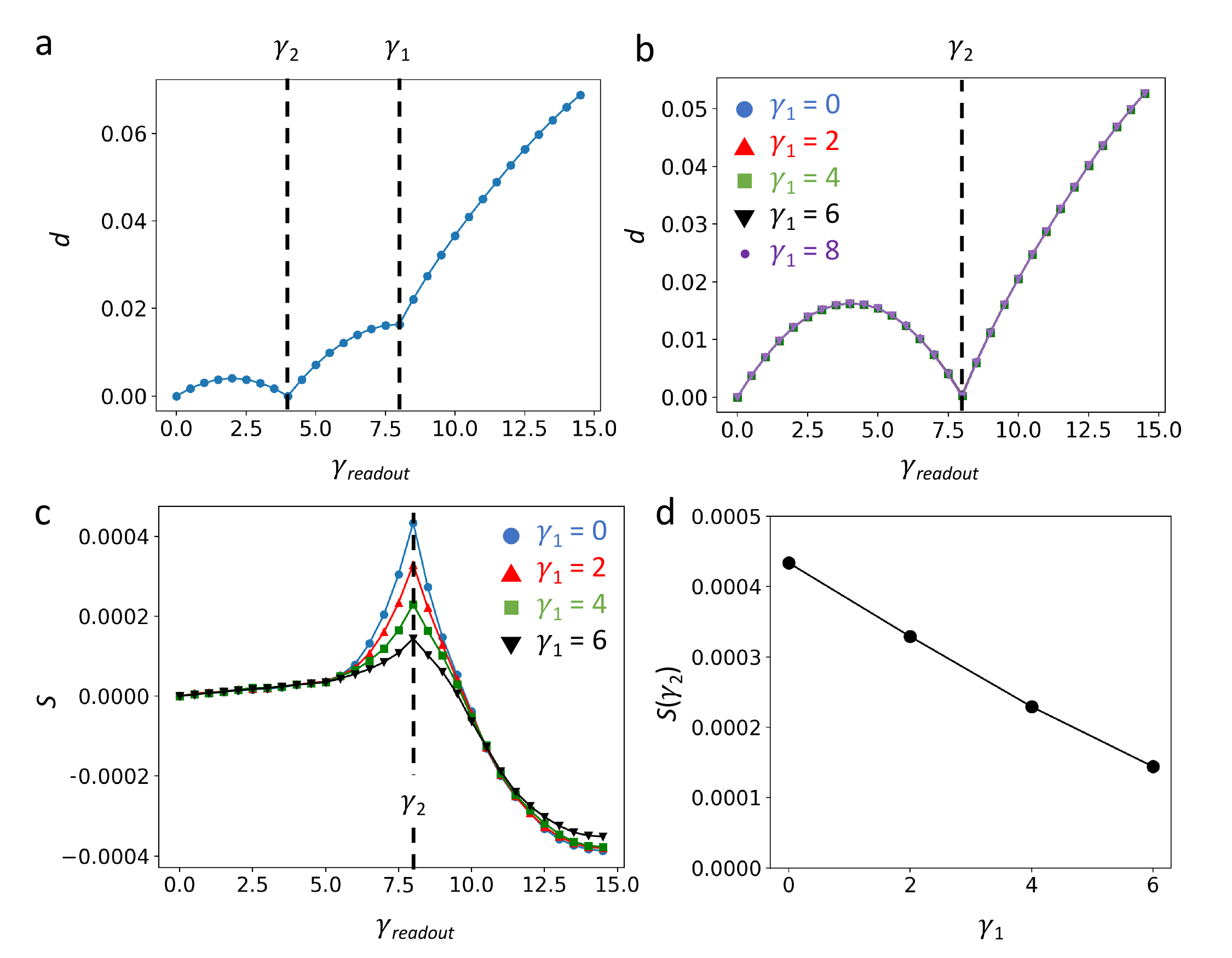}
\caption{
(a) Averaged readout curve for $N=11$ mean field configurations with 
$\gamma_1 > \gamma_2$ in an interacting system with 
interaction strength $A = 0.1$. The curve, which shows two sharp cusps at $\gamma_1$ and 
$\gamma_2$, is nearly indistinguishable from that produced by non-interacting systems
trained in the same way. However,
$d(\gamma_1)$ is precisely zero for non-interacting systems and is small but non-zero
for interacting systems.
(b) Averaged readout curve for $N=11$ mean field
configurations with different values of $\gamma_1$, all less than $\gamma_2$. All systems were trained 
at $\gamma_2 = 8$ and have interaction strength $A = 0.1$. Dashed line shows $\gamma_2$. The
curves lie very nearly on top of one another, so it is
difficult to see the difference between them.
(c) The same data as shown in (b) but with the curve for $d_{\gamma_1 = \gamma_2}$ subtracted off, leaving $S \equiv d - d_{\gamma_1 = \gamma_2}$.  This reveals that there is extra structure in the readout indicating that a memory of the initial (smaller amplitude) input is still encoded in the system. (d) Magnitude of S at $\gamma_2$ as a function
of $\gamma_1$. This magnitude is roughly linear in $\gamma_1$.}
\label{memory}
\end{figure*}

\subsection*{Memories of multiple training inputs}   
As emphasized above, return-point memory gives rise to a hierarchy of memories so that when a larger-amplitude training shear is applied, it erases all memories of previous training with smaller amplitudes.  When we introduce interactions, this is no longer the case; for $\tau >1$, a single shear cycle is by definition not sufficient to bring the system to a periodic state.  It is therefore possible that when a larger strain is applied to a system that has already been trained at a smaller amplitude, there will be a signature left of the initial trained state.  Such behavior was investigated in three  systems showing ``multiple-transient memories'': charge-density-waves ~\cite{Coppersmith97,Povinelli99}, non-Brownian suspensions~\cite{Keim11,Keim13b,Paulsen14} and the park-bench model~\cite{KeimPaulsen2019}.  In those cases, there is a memory of the smaller-amplitude behavior that eventually disappears as more training occurs at the larger amplitudes.  The memory shows up as a cusp in the readout at both training amplitudes.  As training continues, the cusp at the smaller amplitude disappears leaving only the single memory associated with the larger-amplitude input.

We have investigated the possibility of a second memory in our model of interacting hysterons. When we apply our two-amplitude training protocol with a second training pulse that has a smaller amplitude than the first one, $\gamma_2 < \gamma_1$, we see behavior that is similar to the non-interacting case of the Preisach model but with some differences. This is shown in Figure~\ref{memory}a.  As in the Preisach model, there is a sharp cusp at both both $\gamma_1$ and $\gamma_2$.  However, when interactions are introduced, the cusp at $\gamma_1$ occurs at $d > 0$.   

When the second training pulse has a larger amplitude than the first one, $\gamma_2 > \gamma_1$, the system retains a memory of the smaller amplitude input at $\gamma_1$. This would not have been possible in the non-interacting case.
Figure~\ref{memory}b shows the readout for systems trained at different values of $\gamma_1$ and the same value of $\gamma_2$ ($\gamma_2 = 8$) with $\gamma_1 < \gamma_2$ . The curves are nearly the same but not identical; there is a signal that is buried in the small differences between the curves.  This signal can be uncovered by subtracting off the ``background'' (the curve $d_{\gamma_1 = \gamma_2}$, corresponding to $\gamma_1 = \gamma_2 = 8$, which is equivalent to training fully at $\gamma_2$): $S \equiv d - d_{\gamma_1 = \gamma_2}$. As shown in Fig.~\ref{memory}c, $S$  not only has a cusp at $\gamma_2$, but the separate curves have a magnitude at $\gamma_2$ that 
is a linear function of $\gamma_1$ as shown in Fig.~\ref{memory}d. The value of $\gamma_1$ can be determined by the height of $S$.  

\begin{figure}
\centering
\centerline{\includegraphics[width=8.7cm]{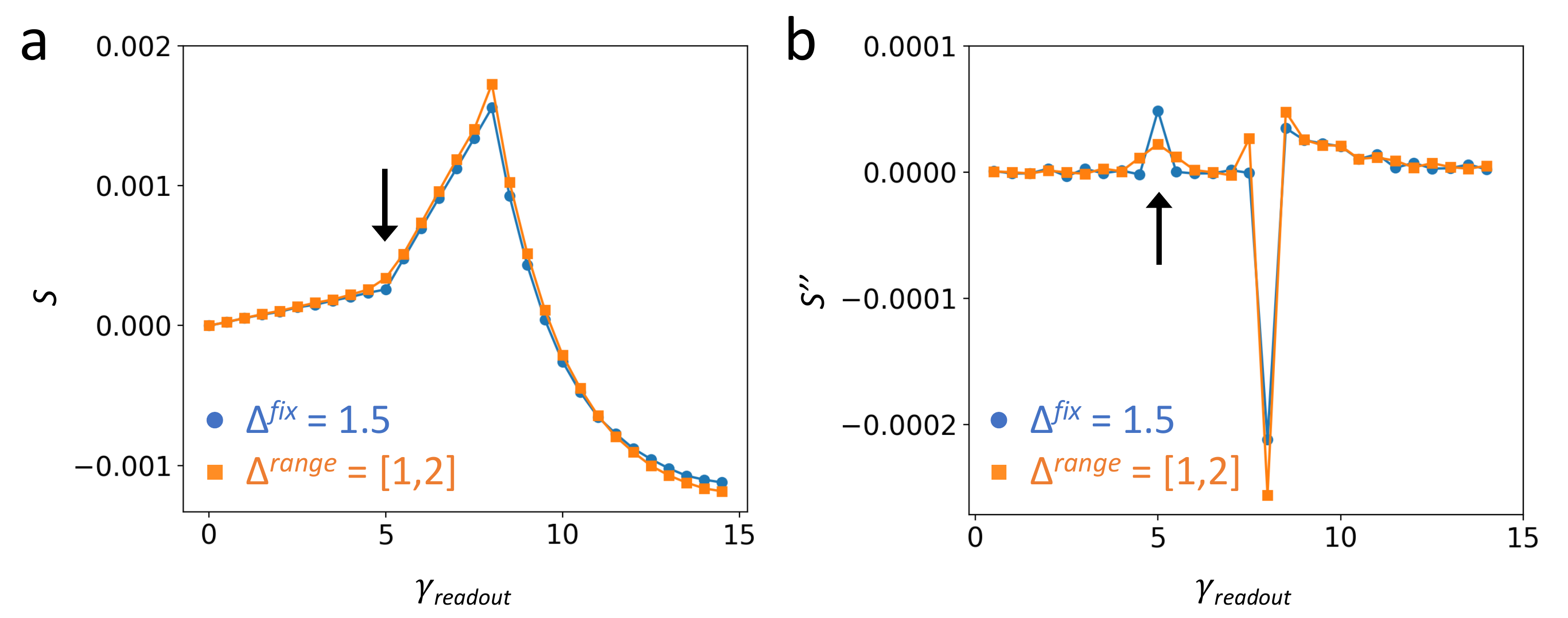}}
\caption{(a) Background-subtracted readout $S$ for 
$\Delta^{fix} = 1.5$ and a uniform distribution $\Delta^{range}$ around $\Delta^{fix} = 1.5$ (so that the interactions are drawn from the range $\pm[1,2]$). All curves shown are for $\gamma_1 = 0$ and $\gamma_2 = 8$. For both types of interaction, there is a change in slope around 
$\gamma_{readout} = 5$, indicated by the black arrow.
(b) $S''$, the second derivative of the background-subtracted readout, versus $\gamma_{readout}$. For $\Delta^{fix}$ interactions,
there is a peak sharply localized around $\gamma_{readout} = 5$, indicated by the black arrow. For $\Delta^{range}$
interactions, the peak is centered on the same value but is broadened.
}
\label{pm}
\end{figure}

When we fix the interactions to be either $+\Delta^{fix}$ or $-\Delta^{fix}$ (so that $\gamma_{i,top}$ jumps by either $0$ or $2\Delta^{fix}$ depending on the state of its neighbors), we 
find that the readout, $S$, has a second cusp structure that appears at ($\gamma_2 - 2 \Delta^{fix}$), as shown in Fig.~\ref{pm}a.  This holds for a range of $\Delta^{fix}$ values. Adding a distribution, $\Delta^{range}$,
around $\Delta^{fix}$ shows a corresponding
broadening of the cusp around ($\gamma_2 - 2 \Delta^{fix}$). This suggests that $S''$, the second derivative of the readout curve with respect to $\gamma_{readout}$, is a measure of the distribution of interactions strengths between the hysterons as shown in Fig.~\ref{pm}b. 

We note that this is different from the double cusps that were seen in the case of multiple transient memories in suspensions and charge-density waves~\cite{Keim11,Keim13b,Paulsen14,Coppersmith97,Povinelli99}.  In that case, the two cusps were determined by the two training amplitudes; here, only one cusp is related to the amplitude $\gamma_2$; the other is determined by the interaction strength $|\Delta^{fix}|$.

\section*{Analysis of two- and three-hysteron systems}

Focusing on small ($N=2$ and $N=3$) systems provides insight
into the periodicity and memory capacity of interacting
hysterons. It has proven useful to describe such systems as directed graphs 
that represent each microstate of the system as a node with arrows showing how the 
system flows to other nodes as shear is applied~\cite{mungan2019networks,Mungan2020StateTransitionGraph}.

\begin{figure}
\centering
\centerline{\includegraphics[width=8.7cm]{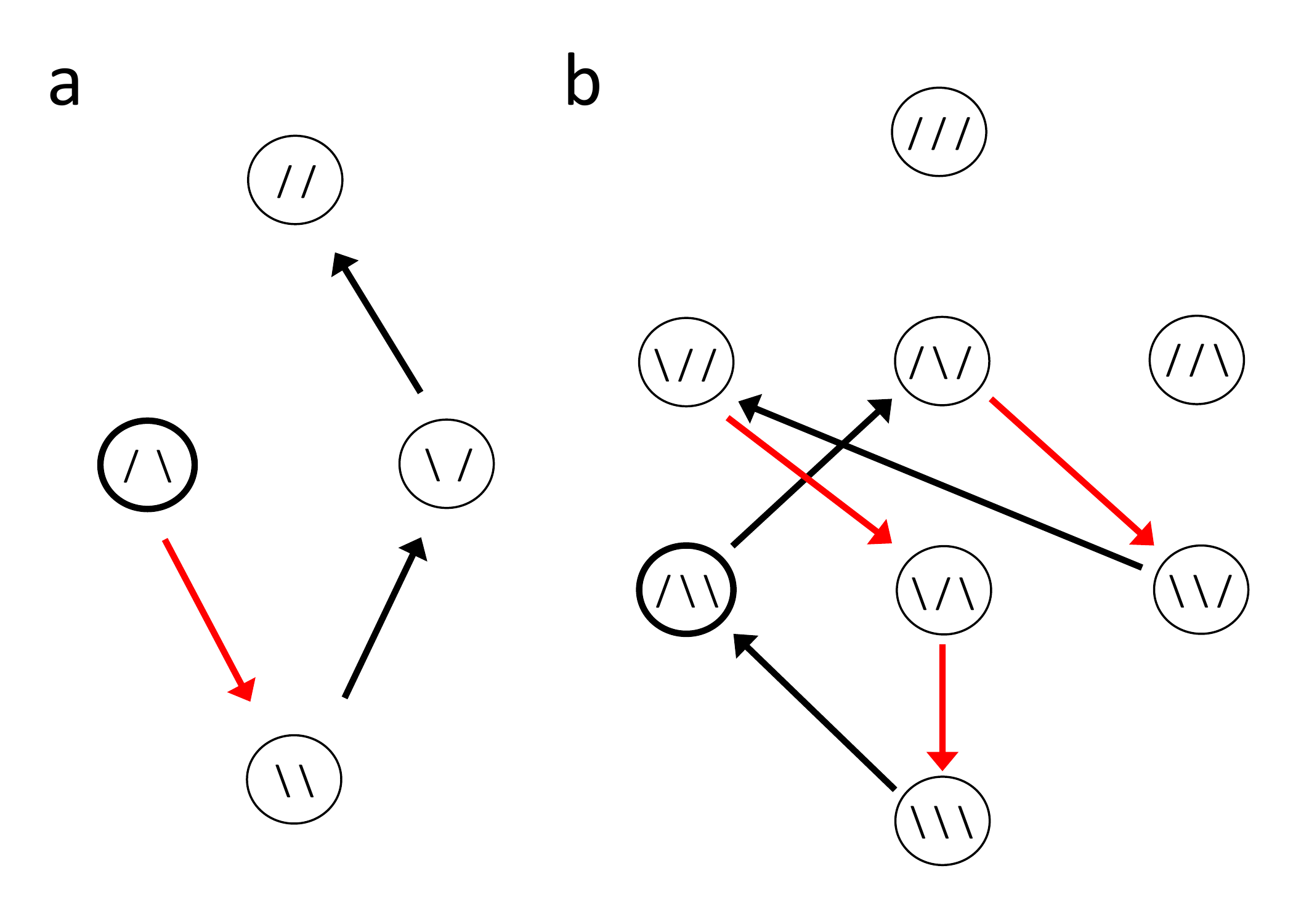}}
\caption{Directed graphs showing $\tau=2$ and $T=2$ behavior. Bolded
microstate represents starting state; black arrows represent 
upward parts of cycle and red arrows represent 
downward part of cycle. (a) Example of $\tau = 2$ in a two-hysteron ($N=2$) system. At end of the first cycle,
system is in the bottom (\textbackslash \ \textbackslash) microstate; 
at end of second cycle
it is in the top (/ /) microstate, where it remains in all
subsequent cycles. (b) Example of $T = 2$ behavior for $N=3$.
}
\label{node}
\end{figure}

The two directed graphs in Fig.~\ref{node} illustrate the possibility
of (a) $\tau = 2$ for $N = 2$ and (b) of $T = 2$ for $N = 3$. 
Such diagrams must follow the rules that correspond to realizable hysteron
configurations; for example, each node must only map to a single node for either increasing or decreasing $\gamma$. Directed graphs make it 
possible to rule out certain behavior, such as $T = 2$ for $N = 2$,
and to enumerate all possible cases of a given behavior.

\subsection*{Exponential decay of $P(T)$}  Although analytic calculations quickly become intractable for large system sizes, an analysis of the case $N=3$ illustrates the nature of the approximately exponential decay of $P(T)$ with $T$ shown in Fig.~\ref{T-and-tau}b.  
Using directed graphs like the one shown in Fig.~\ref{node}b, we can construct all
diagrams corresponding to a $T=2$ configuration 
and assign probabilities to each case. These will depend on the multiplicity
of each diagram and the probability of each arrow connecting
two nodes. The total probability of a given period is obtained by summing over all possible directed graph loops with that period. Although multiplicity of a given loop increases for
larger numbers of microstates visited, there is also
a fractional factor associated with each outgoing arrow, so that higher period loops are suppressed by some small fraction to a high power.  This leads to the generic exponential decrease of $P(T)$ with $T$. 
For simplified model parameters, calculations are analytically tractable and reproduce the simulation results with excellent agreement. 

\subsection*{Memory formation for $N=2$}  
\begin{figure}
\centering
\centerline{\includegraphics[width=8.7cm]{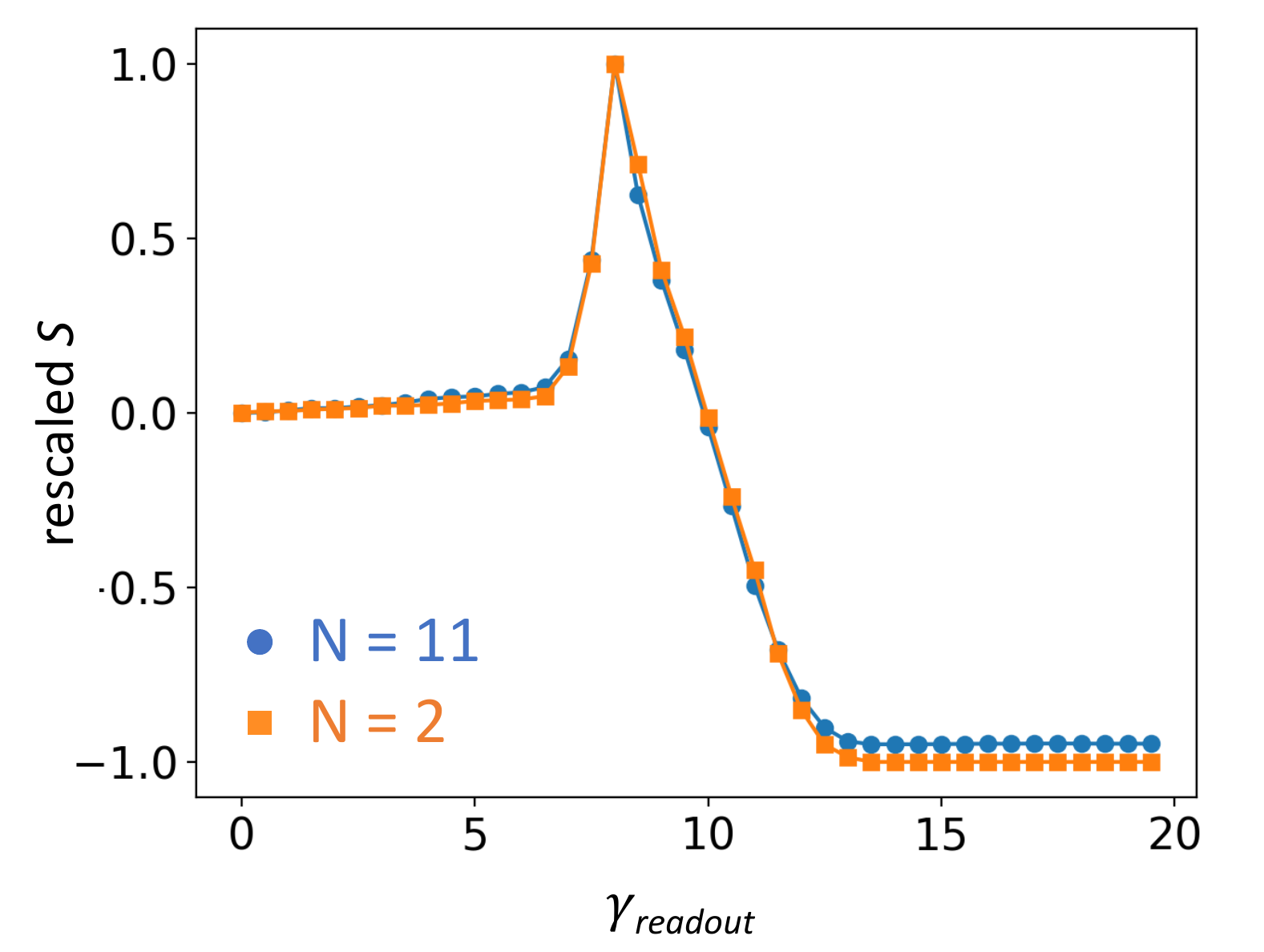}}
\caption{Comparison of background-subtracted readout $S$ for $N=2$ and $N=11$ systems with interaction
strength $A = 0.05$. The curves
agree very well. 
Both curves are for $\gamma_1 = 0$ 
and $\gamma_2 = 8$. They are rescaled to equal $1$ at $\gamma_2$.
}
\label{two-hysteron}
\end{figure}

In terms of memory formation, averaging over many two-hysteron systems ($N=2$) demonstrates several of the key features seen for large systems. We can further simplify the analysis by considering the limit of small interaction strength $\Delta$, which allows us to think of the interactions as perturbations about the non-interacting case. As shown in Fig.~\ref{two-hysteron}, which compares $N=2$ and $N=11$ systems, the background-subtracted readout is nearly independent of $N$
for small interaction strength.

Further analysis of the $N=2$ systems with small interaction strength shows why there is no cusp at $\gamma_{readout} = \gamma_1$.  Enumerating all possible configurations that
contribute to the readout for $\gamma_1 = 0$ and adding
together their contributions, one can understand the shape of $S$. For very small interaction strength,
a single type of configuration dominates. For $\gamma_1$ between $0$ and $\gamma_2$, the details are slightly more complex but the same picture applies. It can be demonstrated in this case that there is no structure at $\gamma_{readout} = \gamma_1$, which shows that this memory is in a different class from those found in suspensions with multiple transient memories. 

\section*{Discussion}

The probabilities of getting $T > 1$ and $\tau > 1$ 
and the exponential decay of $P(T)$ with $T$,
as observed in~\cite{lavrentovich2017period}, are successes of this model. 
In cyclically sheared jammed packings, we expect systems closer to jamming (that is,
at lower pressure) to have longer-range interactions, effectively coupling a larger
number of rearranging regions. The increase in $P(T>1)$ with system size in our systems of
hysterons is therefore consistent with the general trend seen in 
\cite{lavrentovich2017period}, where $P(T>1)$ increases with decreasing pressure.
The effect of relative location of the interacting regions has been studied in the paper by Keim and Paulsen~\cite{KeimPaulsen2020}.

We see two distinct types of memory in Fig.~\ref{memory}c.
As in non-interacting systems of hysterons, $\gamma_2$ is stored locally in the location of the cusp of the readout signal. There is no local signature such as a cusp in the response associated with the amplitude of $\gamma_1$ if $\gamma_1 < \gamma_2$, in accord with simulations of jammed systems~\cite{Adhikari18}.  (This is in contrast to the cusp-like second memory that was found in non-Brownian suspensions and charge-density waves~\cite{Keim11,Keim13b,Paulsen14,Coppersmith97,Povinelli99}.) However, there is still a memory of $\gamma_1$ that is embedded in the \textit{magnitude} of the 
readout; by calibrating the response
it is possible to determine the exact value of $\gamma_1$. This is a new form of memory.
Our results suggest that 
we can reinterpret readout curves 
from simulations of jammed systems for different numbers 
of training cycles~\cite{fiocco2014encoding} likewise
as a memory of the number of training cycles applied to the system.

It is especially exciting that this new memory provides a way to measure the strength of the interactions present in 
an interacting system. We showed that for two delta-functions (at
$\pm \Delta^{fix}$) we can read out the interaction strength directly from the readout signal. This survives to 
broader distributions. Thus, the shape of the readout signal $S$ provides 
information about typical interaction strengths in the system. This could provide a new, possibly unique, method to determine this crucial parameter in real systems. A natural next step is to 
apply this training protocol to simulations of a jammed system. If our interacting model successfully captures 
the memory behavior of such systems, it could provide a simple method
for measuring the distribution of interaction strengths between rearrangements. 

\section{Methods}

For simulations of the mean-field model, we report the results for two types of situations: (i) The interaction strengths $\Delta^{nn}_{i,top}$ and 
$\Delta^{nn}_{i,bot}$ for hysteron $i$ are chosen
uniformly from  
$[-A*(\gamma^0_{i,top}-\gamma^0_{i,bot}),+A*(\gamma^0_{i,top}-\gamma^0_{i,bot})]$ so that the interactions are scaled by the length of the hysteron they affect. That is, longer hysterons are more likely to have their flipping strains shifted by larger values. We choose $A$ between $0$ and $0.5$. The distribution of
initial hysteron lengths 
($\gamma^0_{i,top}-\gamma^0_{i,bot}$) and  midpoints $(\gamma^0_{i,top}+\gamma^0_{i,bot})/2$ are drawn randomly from between $[0,20]$ and $[-25,25]$ respectively.
(ii) We fix the values of $\Delta^{nn}_{i,top}$ 
to be 
$\pm \Delta^{fix}$ so that $\gamma_{i,top} = \gamma^0_{i,top} \pm \Delta^{fix}$ and likewise for $\gamma_{i,bot}$. In this case the shifts are
not scaled by the length of the hysteron they affect. We choose the distribution of
($\gamma^0_{i,top}-\gamma^0_{i,bot}$) and $(\gamma^0_{i,top}+\gamma^0_{i,bot})/2$ to be drawn randomly from between $[5,20]$ and $[-25,25]$ respectively. 
For both (i) and (ii), the results are averaged over between $10^4$ and $10^7$ independent configurations.

\section{Acknowledgements}
We are grateful to N. Keim and J. Paulsen for sharing their manuscript prior to publication.  We thank them and P. Littlewood, M. Mungan, S. Sastry, A.K. Sood, M. van Hecke and Z. Zeravcic for many discussion about memory formation in matter and M. Lavrentovich for also sending data used in the inset of Fig~\ref{T-and-tau}b. This work was supported by the US Department of Energy, Office of Science, Basic Energy Sciences, under Grant DE-SC0020972 (for theoretical analysis and model development) and by the Simons Foundation for the collaboration “Cracking the Glass Problem” Award 348125 (for simulations). C.W.L. was supported by a National Science Foundation Graduate Research Fellowship under Grant DGE-1746045.

\bibliography{main}

\end{document}